# Bidirectional magnetization switching of a ferrimagnetic insulator with monochiral molecules

*Wei-Hsiang Liao*[a‡], *Joshua S. Webb*[b,e‡], *Yu-Hui Xue*[a], *Yao Zhang*[c], *Yu-Ying Chang*[a],

*Muhammad Ali Hashmi*[d,e], *Patricia A. Hunt*[d,e], *Simon Granville*[c,e]*, *Yu-Cheng Shao*[f]*,

*Muhammad Hanif*[b,e]*, and *Hua-Shu Hsu*[a]*

[a]Department of Applied Physics, National Pingtung University, No. 4-18 Minsheng Rd., Pingtung 900391, Taiwan

[b]School of Chemical Sciences, The University of Auckland, Private Bag 92019, Auckland 1142, New Zealand

[c]Robinson Research Institute, Victoria University of Wellington. P.O. Box 33–436, Lower Hutt 5046, New Zealand

[d]School of Chemical and Physical Sciences, Victoria University of Wellington, Wellington, New Zealand

[e]MacDiarmid Institute for Advanced Materials and Nanotechnology, New Zealand

[f]National Synchrotron Radiation Research Center, Hsinchu, 300092, Taiwan

*Correspondence and requests for materials should be addressed to Simon Granville (email: simon.granville@vuw.ac.nz), Yu-Cheng Shao (email: shao.yc@nsrrc.org.tw), Muhammad Hanif (email: m.hanif@auckland.ac.nz) and Hua-Shu Hsu (email: hshsu@mail.nptu.edu.tw).

**ABSTRACT**

Recent studies have demonstrated that magnetization switching in ferromagnets can be achieved through adsorbing chiral molecules on the surface without the need for current or external magnetic fields, offering a low-power mechanism for applications in spintronic devices. Opposite chirality molecules cause opposite direction reversals of magnetization through the chiral-induced spin selectivity (CISS) mechanism. In this study, we demonstrate bidirectional magnetization switching in thin films of ferrimagnetic insulator TmIG using a *single* chirality molecule – a Cu metallopolymer of D-leucine.

Through UV-VIS circular dichroism and X-ray absorption spectroscopy, we determined that switching between different magnetic orientations is associated with interactions of the D-leucine with the two distinct sublattices of the Fe ions in the TmIG, at octahedral and tetrahedral sites. Our study demonstrates the unexpected versatility of the CISS mechanism for magnetization switching in ferrimagnets using single chirality materials, thereby expanding the potential applications of chiral molecule adsorption-induced magnetization flipping.

## INTRODUTION

Recent studies have successfully demonstrated magnetization switching in ferromagnets by adsorbing chiral molecules, without the need for external magnetic fields or current injection[1-4]. These findings introduce a simple, low-power route for controlling magnetization under ambient conditions, with the magnetic orientation governed by the handedness of the chiral molecules. Two main effects are widely considered to underlie this phenomenon. First, magnetism induced by the proximity of adsorbed chiral molecules arises from spin-selective electron transfer, leading to spin polarization in the underlying metallic layer[1,2,5]. Second, the chiral-induced spin selectivity (CISS) effect dictates that electron spins align either parallel or anti-parallel to their velocity, depending on the molecular chirality[6-9], producing spin-polarized currents capable of influencing magnetic order.

Building on these mechanisms, magnetization switching has been demonstrated in perpendicularly magnetized ferromagnetic layers coated with gold via molecular adsorption of chiral molecules[1]. These advances prompt a forward-looking question: Is it possible for chiral molecules of a single handedness to induce magnetic moment reversal in opposite directions across different magnetic materials?

To address this question, we selected ferrimagnetic thulium iron garnet (TmIG) as the substrate material for direct adsorption of chiral molecules. TmIG was chosen due to its ferrimagnetic nature, characterized by two opposing magnetic sublattices that result in a net magnetic moment, making it suitable for spin-related interactions with chiral species. In the ferrimagnetic iron garnets, the magnetic moments of the $Fe^{3+}$ ions in the octahedral ($O_h$) sites are antiparallel to those in the tetrahedral ($T_d$) sites. The magnetization direction is primarily dominated by the $T_d$ sites, which are occupied by $Fe^{3+}$ ions with stronger exchange interactions compared to those in the octahedral sites. The magnetic moments in the tetrahedral and octahedral sites are antiferromagnetically coupled, but the $T_d$ $Fe^{3+}$ ions typically have a greater influence on the net magnetization direction due to their stronger contribution to the overall exchange field.[10,11]

TmIG is widely regarded as a promising material for spintronic device applications, owing to its compatibility with thin-film growth and inherent perpendicular magnetic anisotropy.[12] In addition, electrical current-induced magnetization switching has been successfully demonstrated in TmIG-based systems, underscoring its potential for dynamic spin control.[13-15] These characteristics make TmIG a suitable platform for our investigation of interfacial interactions between magnetism and chiral materials. In the selection of chiral molecules, we chose Cu complexes of L- and D-leucine. For magnetic exploration, we opted for UV-Vis magnetic circular dichroism (MCD) spectroscopy, which is capable of simultaneously indicating magnetic orientation and spin-polarized energy bands and has been proven to be a powerful tool in studying the band structure of spin-polarized materials.[16-19]Additionally, we employed X-ray absorption spectroscopy (XAS) to examine changes in the electronic structure of Fe in TmIG after adsorption of these chiral molecules.

Our main experimental result is that magnetization switching of TmIG in both directions is achieved using a monochiral Cu complex of D-leucine. Surprisingly, the opposite chirality L-leucine does not cause magnetization switching in either direction. In addition, there was no change in magnetization when L- and D-leucine without Cu were adsorbed on TmIG.

## RESULTS AND DISCUSSION

### Synthesis and characterization of chiral Cu compounds

Recently, Naaman and co-workers reported that chiral metal–organic materials demonstrated the CISS effect.[20-22] Motivated by these findings, we prepared chiral copper coordination polymers using enantiopure L- and D-leucine. Amino acids such as leucine are cheap, readily available, and excellent feedstocks of chiral ligands, offering scope for scaling up of materials. Metal coordination occurs through the deprotonated carboxylic acid moiety and the primary amine of the leucine enantiomers.[23] This results in the formation of neutral complexes. The Cu complexes are relatively less soluble in most organic solvents and water. Metal coordination was indicated by the appearance of N-H stretches around 3250 $cm^{-1}$ and the loss of $NH_3$ stretches around 3000 $cm^{-1}$ in FTIR spectra. The identity and purity of Cu compounds were confirmed by microanalysis, where the experimental elemental composition matches very well with the calculated values for the Cu(II) complexes of L- and D-leucine. The Cu(II) complexes of l- and d-leucine have been optimised and confirmed as minima at the M06-2X-PCM(water)/6-311+G(d,p) level of theory (see computational methods for details). A range of intramolecular H-bonding interactions between ligands stabilise the metal-organic structures (Fig. 1).

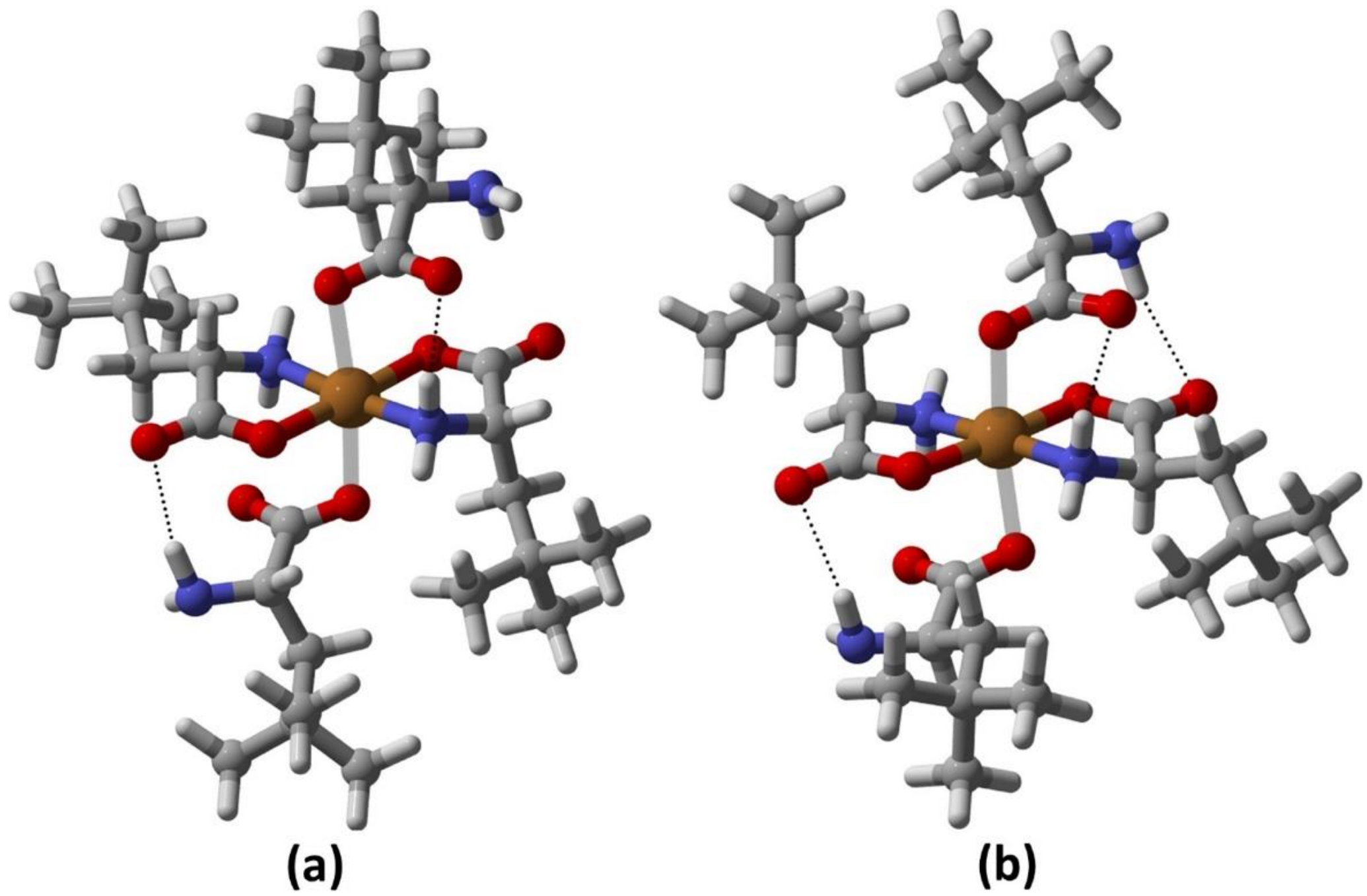

**Figure 1.** Optimized geometries of Cu(II) complexes of D-leucine (a) and L-leucine (b) calculated at M06-2X-PCM(water)/6-311+G(d,p) level of theory.

**Solution Circular Dichroism (CD) spectroscopy**

We used CD spectroscopy to confirm the chirality of the Cu(II) complexes of L-leucine and D-leucine. In Figure 2(a), we present the CD spectra of L-leucine and D-leucine with and without copper. The opposite CD absorption spectra of the copper complexes of the two enantiomers indicate that the metal-organic structures maintain the chirality of the starting amino acids. However, due to the spin-orbit interaction, the CD spectra underwent significant changes after Cu coordination to the leucine. The original single Gaussian peak around 6 eV transformed into a differential function type graph with a positive peak near 4.5 eV and a negative trough at 6 eV. The Electronic CD (ECD) spectra of d- and l-leucine complexes have been calculated at the CAM-B3LYP-D3BJ-PCM(water)/6-311+G(d,p) level on the M06-2X-PCM(water)/6-311+G(d,p) optimised structures and compared to the experimental spectra, giving a qualitative visual match (Fig. S2 and S3).

First, the MCD spectrum of the TmIG films is measured without chiral molecules. The film was initially magnetized in the out-of-plane (+z) direction with a magnetic field of 0.4T, then the field removed to set the film in the remanent state. In Figure 2(b), the MCD spectrum of TmIG is shown after the applied magnetic field has been removed (blue line). Multiple characteristic peaks are observed in the MCD spectrum, which is typical for ferrimagnetic oxides with magnetic sublattices[15,25,26]. With the applied magnetic field removed, the TmIG film still maintains its CD signal, indicating an easy axis perpendicular to the film surface due to the perpendicular magnetic anisotropy. By applying a reverse magnetic field of 0.4T, the sign of the CD spectrum is reversed (red line), showing we can reverse the magnetization direction. Therefore, in subsequent experiments, we initially set the TmIG in +z (magnetized up) and -z (magnetized down) remanent states before adsorbing chiral molecules, to investigate the subsequent CD changes from those remanent states. Notably, neither the Cu complex of L-leucine nor the two enantiomers of leucine without Cu exhibited the flipping in TmIG magnetization. The explanation for this unexpected behavior is addressed in a later section. Therefore, we focused further studies on the Cu complex of D-leucine.

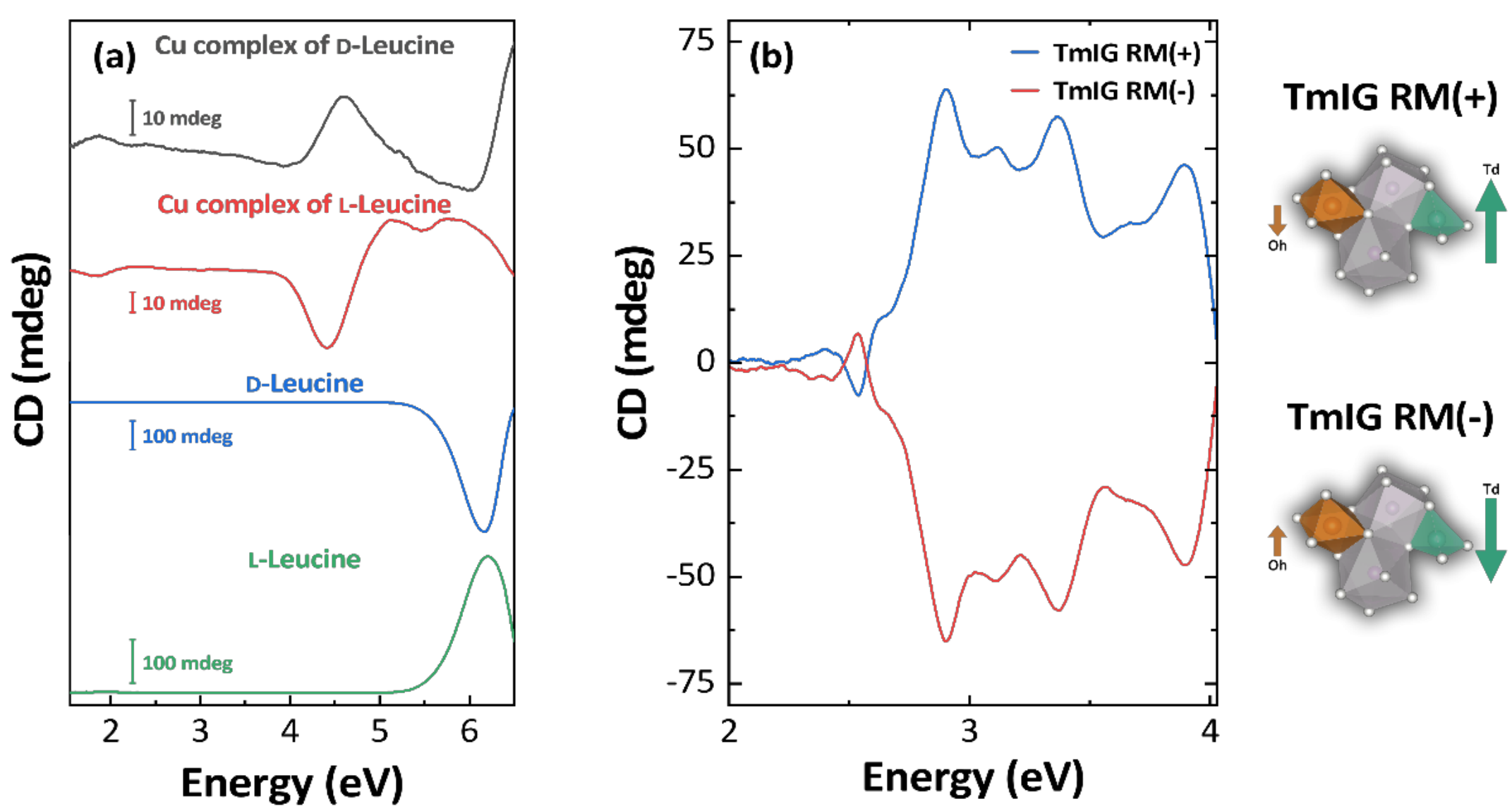

**Figure 2.** (a) The CD spectra of samples of L-leucine, D-leucine, and their Cu complexes, respectively, in solutions. (b) The MCD spectrum of TmIG in its remanent magnetic states after saturation in applied out-of-plane magnetic fields of 0.4T. The magnetic moments of the $Fe^{3+}$ ions

in the $O_h$ sites are antiparallel to those in the $T_d$ sites. The magnetization direction is primarily dominated by the $T_d$ sites.

**Bidirectional Magnetization Reversal in TmIG**

In Figure 3(a), we observe the CD spectra of TmIG before and after the adsorption of the Cu complex of D-leucine, when the TmIG remnant magnetic moment was initialized along the +z direction (RM+). Notably, the MCD spectrum after adsorption closely resembles the RM- signal in Figure 2(b), which shows the magnetization is reversaled to the –z direction after adsorption of the Cu complex of D-leucine, confirming the chiral-induced magnetization switching of TmIG using the Cu complex of D-leucine. The CD spectrum remains unchanged from the initialized +z direction when the opposite chirality L-leucine is adsorbed, for either the pure or Cu complex molecules (refer to Supplementary Information S1). Interestingly, the CD spectrum of TmIG is also unchanged when D-leucine molecules without Cu are adsorbed. This result indicates that the presence of Cu in the D-leucine plays a crucial role in facilitating the magnetization reversal process.

In Figure 3(b) we show the results of the same experiment except with the TmIG magnetization initialized along the -z direction (RM-). Unexpectedly, magnetization switching from –z to +z is seen in the MCD spectrum when adsorbing D-leucine chiral molecules containing Cu – the same chirality molecule that demonstrates magnetization switching from +z to -z. Our results therefore show *bi*directional magnetization switching in ferrimagnet TmIG with a *single* chirality molecule. This differs from the situation for ferromagnetic materials and chiral molecules, where the flipping of the magnetic moment from pointing along the +z direction to the -z direction, or vice versa, requires molecules of different chirality. At first glance, this observation appears to deviate from the conventional model of CISS-induced magnetization switching, which predicts that molecules of opposite chirality should induce opposite effects due to their generation of spin-polarized carriers with opposite spin orientations.[1] However, in our experiment, a single chirality molecule is able to drive magnetization reversal in both directions depending on the initial magnetic state. While this behavior may seem inconsistent with the CISS framework, it may still be reconciled when considering the ferrimagnetic nature of TmIG, which comprises two magnetic sublattices with antiparallel spin alignments. This point is discussed in a later section, where we propose a mechanism in which spin-selective interactions between the single chirality molecule and the distinct sublattices enable bidirectional magnetization switching.

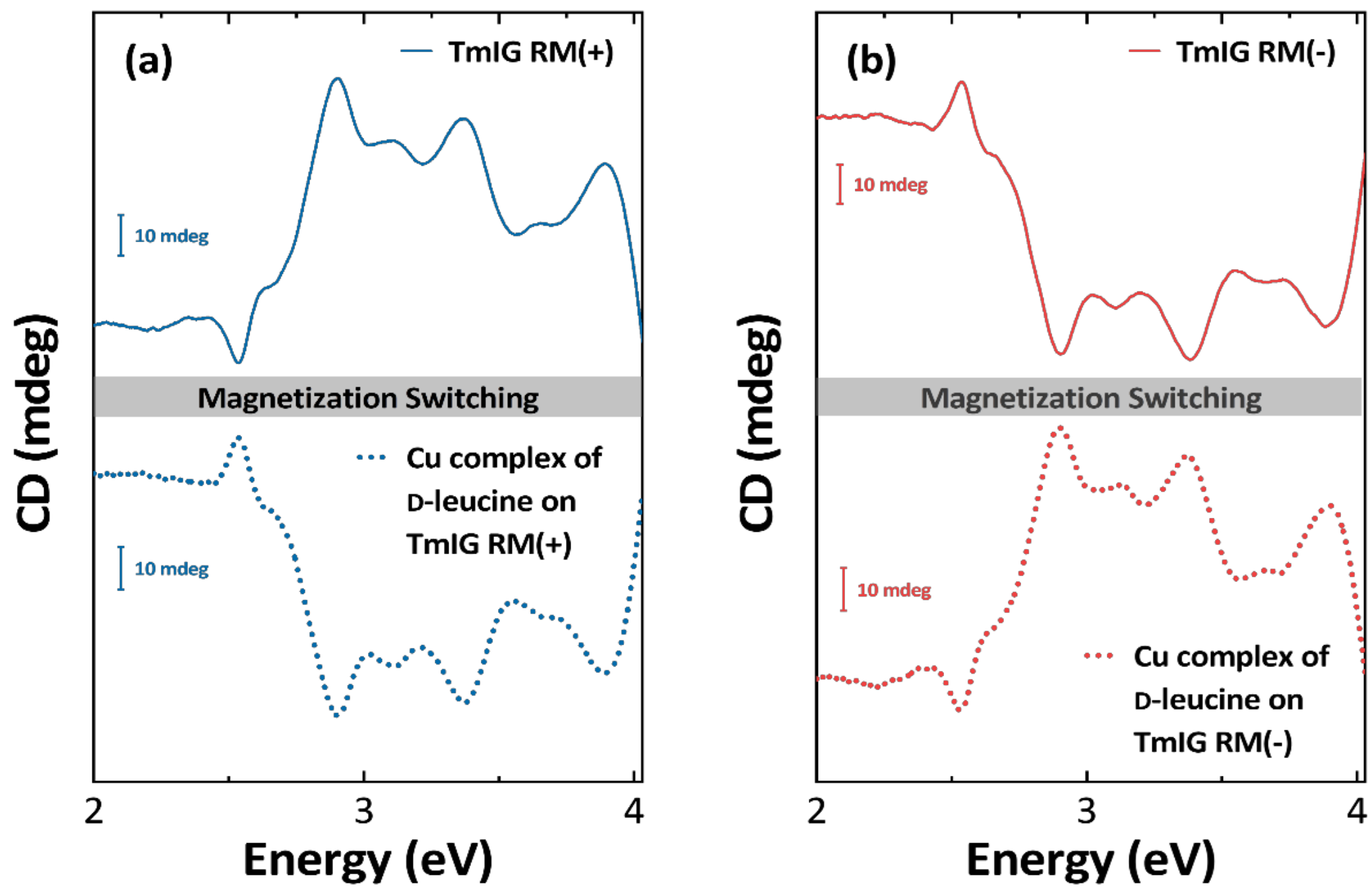

**Figure 3.** (a) CD spectra of TmIG before and after adsorption of Cu complex of D-leucine, when the remanent magnetic state is RM(+). A magnetization reversal to the –z direction is observed in the CD spectra of TmIG after the adsorption of Cu complex of D-leucine. (b) Additionally, Cu complex of D-leucine can also induce the reversal of the magnetic moment from the initial RM(-) state to the +z direction.

**Probing the Influence of Chiral Molecular Adsorption on Fe K-Edge XANES**

To explore the possible mechanism by which molecules of a single chirality can induce magnetization reversal from both RM(+) and RM(-) initialized states, we conducted X-ray Absorption Near Edge Structure (XANES) measurements at the Fe K-edge to investigate the electronic structure modifications induced by the chiral molecules. Figure 4(a) presents the XANES results of four samples: TmIG films in both the RM(+) and RM(-) initialized states before and after adsorption of the Cu complex of D-leucine. Notably, no significant spectral differences were observed between the RM(+) and RM(-) samples, where the only difference is their opposite magnetization directions. However, the spectra of RM(+) and RM(-) samples after adsorption of the D-leucine exhibit subtle yet distinct differences. Compared to the original RM(+) and RM(-) samples, after molecular adsorption, the RM(+) sample displayed lower absorption in the pre-edge region and the RM(-) sample higher absorption (Fig 4(a) inset).

In principle, the K-edge XANES in 3d transition metals such as Fe is primarily dominated by excitations from 1s to 4p electronic states, while Fe 3d–4p hybridization can introduce pre-edge features containing 3d electronic information[27,28]. However, in TmIG, the pre-edge features of the Fe K-edge primarily originate from Fe(3d)–O(2p) hybridization. Furthermore, since TmIG consists of two magnetic sublattices, Fe($T_d$) and Fe($O_h$), the simulated spectra in Figure 4(b) indicate that the Fe K-edge XANES of TmIG includes contributions from $Fe^{3+}$ at both $T_d$ and $O_h$ sites. Specifically, the pre-edge intensity is higher for $T_d$ and lower for $O_h$. This is because the $T_d$ site has lower symmetry, leading to stronger Fe(3d)-O(2p) hybridization, which enhances the pre-edge absorption signal. In contrast, the $O_h$ site exhibits higher symmetry, resulting in weaker hybridization and thus lower pre-edge absorption intensity. The changes in the XANES spectra of RM(+) and RM(-) samples after molecular adsorption may be associated with the differential effects of molecular adsorption on the Fe($O_h$) and Fe($T_d$) magnetic sublattices. This observation provides further insights for future investigations.

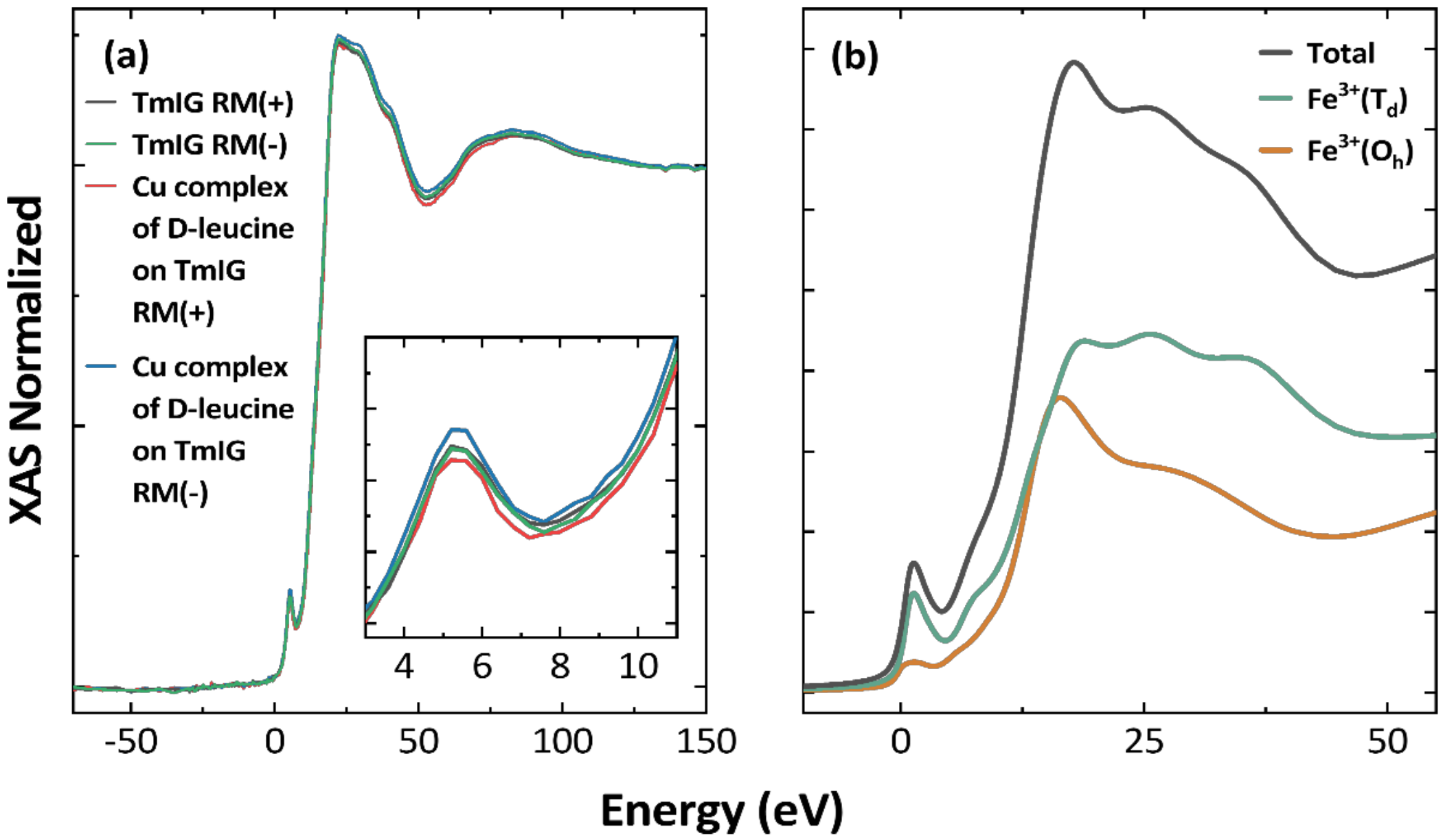

**Figure 4.** (a) XANES spectra for four samples: RM(+) and RM(-) initialized states before and after adsorption of the Cu complex of D-leucine, where magnetization reversal occurs upon adsorption. The inset shows a magnified view of the pre-edge region. (b) Simulated Fe K-edge XANES spectra, illustrating the absorption features of $Fe^{3+}$ at the $T_d$ and $O_h$ sites. The pre-edge absorption peak intensity for the $T_d$ site is higher than for the $O_h$ site.

**Understanding the Mechanism of Magnetization Reversal after Cu complex of d-leucine absorption**

Since CD spectroscopy can resolve spin-polarized electronic structures, we further analyze the changes in the CD spectrum to understand how the single-handed chiral Cu complex of D-leucine can induce magnetization reversal in TmIG from either remanent magnetic direction, while L-leucine cannot achieve switching from either direction.

First, the magnitude of the CD signals before and after magnetization reversal from Figure 3(a) and 3(b) were overlaid in 5(a) and 5(b). Plotting the magnitude of the CD signals removes the different sign of the spectra between initial and magnetization reversed states, allowing for a clearer visualization of the CD variations induced by chiral molecular adsorption.

Notably, in Figure 5(a), after Cu complex of D-leucine adsorption on TmIG in the RM(+) state, the features in the CD spectrum above 3.3 eV are significantly weaker. This energy range is primarily dominated by transitions from the occupied O(2p) states to the unoccupied Fe($T_d$) states. In Figure 5(b), by contrast, when comparing the CD intensity after Cu complex of D-leucine adsorption on TmIG starting in the oppositely magnetized RM(-) state, the most prominent changes occur below 3.3 eV, in the energy range attributed to transitions from the occupied O(2p) states to the unoccupied Fe($O_h$) states. Hence, the initial remanent magnetization direction determines which of two different ranges of energy transitions is affected by adsorbing a single chirality Cu complex of D-leucine on TmIG.  These two energy ranges involve the two different Fe sublattices in TmIG.  The interaction between the adsorbed molecules and the two anti-ferromagnetically aligned sublattices is the key to understanding the mechanism for the monochiral switching in TmIG and provides a distinct difference to the case of CISS switching in ferromagnets, where all magnetic moments are oriented in the same way.

A schematic of the spin-polarized density of states of TmIG is shown in Figure 5(e). Upon adsorption of the Cu complex of D-leucine onto TmIG, we observe a reversal of the net magnetization, primarily originating from Fe ions at $T_d$ sites. Concurrently, the MCD signal corresponding to O 2p → Fe($T_d$) 3d transitions is noticeably reduced, indicating a modification of the spin-polarized unoccupied states at these sites, as displayed in the Figure 5(f). This behavior is attributed to the CISS effect, wherein the chiral molecular layer interacts spin-selectively with the surface states of TmIG. Specifically, the Cu– D-leucine complex couples preferentially to the originally spin-up Fe($T_d$) states due to its handedness, as shown in the Figure 5(c). This interaction perturbs the electronic and magnetic environment of the Fe($T_d$) sublattice, leading to a reorientation of its magnetic moment. This spin-filtering effect introduces an imbalance in the sublattice spin populations by selectively modifying the Fe($T_d$) contribution. As a result, the ferrimagnetic order is disrupted, and the net magnetization is reversed. The accompanying reduction in MCD intensity further supports this scenario, reflecting a diminished contribution from Fe($T_d$) to the MCD signal after adsorption.

In the case of the RM(–) sample, the net magnetization prior to molecular adsorption is still dominated by Fe ions at $T_d$ sites, which are now aligned in the spin-down direction. Due to the chiral-induced CISS effect, the Cu complex of D-leucine preferentially interacts with spin-up states.

In this reversed spin configuration, it is the Fe($O_h$) sublattice that hosts the spin-up orientation and thus becomes the primary site of interaction. This selective coupling perturbs the spin polarization of the Fe($O_h$) sublattice, leading to a reversal of its magnetic contribution. Consequently, the delicate balance between the Fe($O_h$) and Fe($T_d$) sublattices is disrupted, resulting in a net reversal of the total magnetization. The MCD spectrum supports this interpretation by showing a noticeable suppression in the Fe($O_h$)-related features, reflecting a diminished contribution from Fe($O_h$) to the MCD signal after adsorption (<3.3 eV). This observation further confirms that the direction of magnetization switching is not determined by the initial magnetization polarity, but rather by the spin-dependent coupling between the chiral molecule and the spin-resolved sublattice states of the ferrimagnet, as governed by the CISS mechanism.

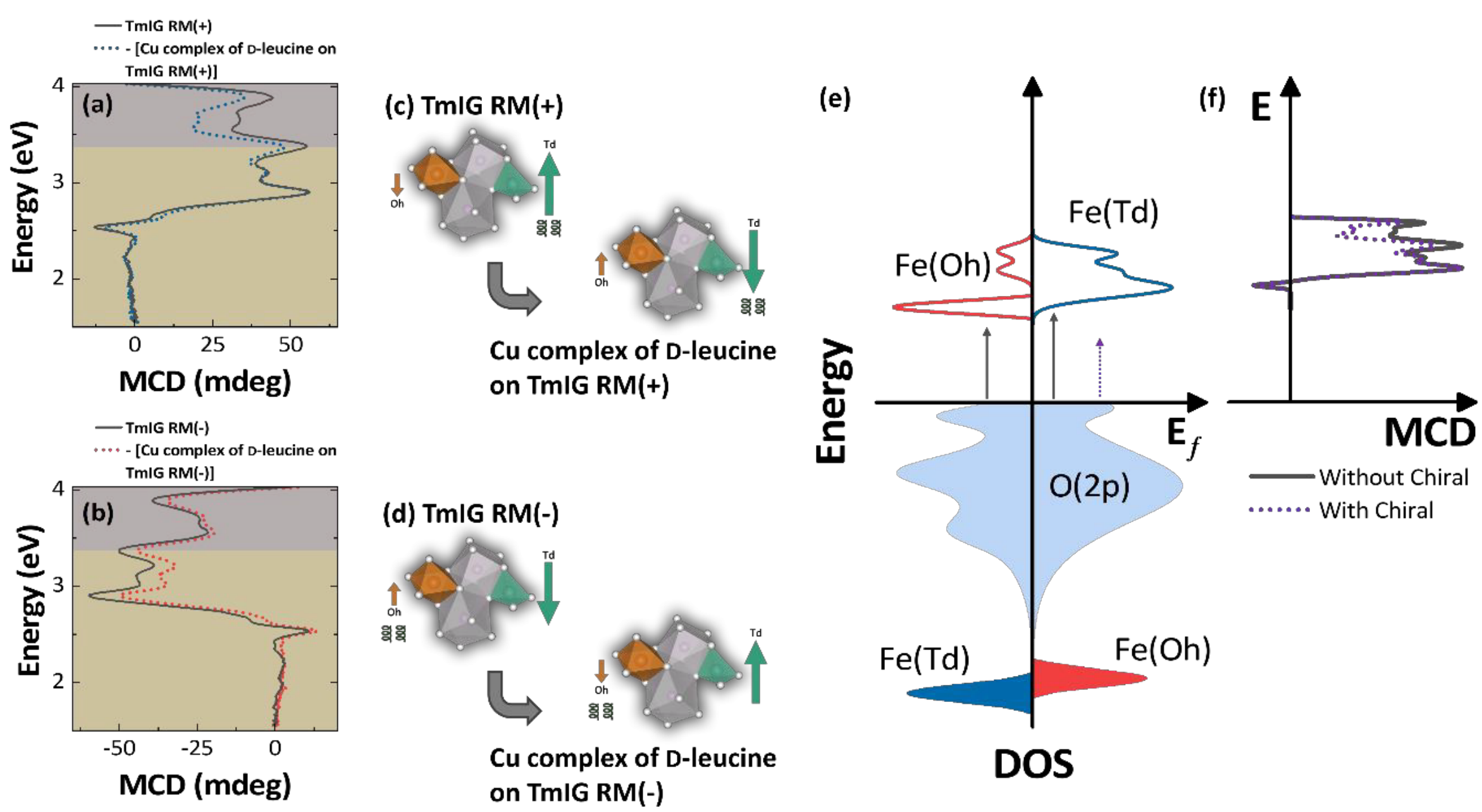

**Figure 5.** The CD signals from Figures 3(a) and (b), where the magnetization moment was reversed due to the adsorption of D-leucine. The CD spectra are plotted together without the sign reversal that takes place when adding the D-leucine to make it easier to see how the features change. The spectra are plotted starting from the (a) RM(+) and (b) RM(-) remanent states. (c) The reversal of the net magnetic moment after RM(+) adsorption of Cu complex of D-leucine, based on the spin-dependent electronic band structure. The interaction between Cu complex of D-leucine and Fe($T_d$) magnetic sublattices explains the observed CD change in (a). (d)The reversal of the net

magnetic moment after RM(-) adsorption of Cu complex of D-leucine, also illustrating the magnetic sublattice selectivity. In this case, the Cu complex of D-leucine interacts with the Fe($O_h$) sublattice. (e) A schematic of the spin-dependent electronic structure,[14] and (f) the corresponding CD spectra.

To summarize this discussion, we have identified from the CD and XANES spectroscopy that the Cu complex of D-leucine primarily interacts with the spin-up magnetic sublattice in TmIG. Specifically, in the RM(+) state, the Cu complex of D-leucine interacts with Fe($T_d$), whereas in the RM(-) state, it interacts with Fe($O_h$). This finding explains why a molecule with a single chiral orientation can induce magnetization reversal in both RM(+) and RM(-) states. This phenomenon arises because the Cu complex of D-leucine follows CISS, and TmIG, as a ferrimagnet, consists of two magnetic sublattices with opposite spin orientations. Consequently, regardless of whether the remanent magnetization is along +z or -z, Cu-doped D-leucine maintains its spin selectivity and specifically only interacts with one of the magnetic sublattices, leading to the reversal of the magnetic moments within that specific sublattice. Furthermore, due to the intrinsic antiferromagnetic coupling between the two magnetic sublattices in TmIG, when one magnetic sublattice undergoes magnetization reversal, the net magnetization of the entire system is also reversed.

Our results have important implications for CISS applied to magnetic materials. Non-ferromagnetic materials, e.g. ferrimagnets, antiferromagnets, and materials with non-collinear magnetic moment alignments, may have complex interactions with chiral molecules, due to the presence of multiple magnetic sublattices and competing exchange interactions. Introducing magnetic materials with these types of magnetic order to CISS studies may provide many new and interesting ways to manipulate magnetization and spin polarization. Our results on ferrimagnetic TmIG show that a single-handed chiral molecule can be used to reverse the magnetization. This also makes applications based on molecular-induced magnetization switching more feasible, by removing the complexity of needing two different enantiomers to make the switch reversible.

## CONCLUSION

In summary, our study demonstrates that magnetization switching in ferrimagnetic TmIG can be directly achieved in either orientation through the adsorption of a single-handed chiral molecular species, specifically Cu-containing D-leucine. CD and XANES spectroscopies reveal differences in the TmIG electronic states that are affected by adsorption of this molecule, depending on the initial remanent magnetization direction of the TmIG. In contrast to ferromagnets, which have only one magnetic lattice, this bidirectional switching is due to the chiral molecules interacting with a different Fe sublattice in TmIG, depending on the initial magnetization direction. The effect of the chiral molecule on the chosen Fe sublattice, combined with the antiparallel coupling of that sublattice with the other one, results in magnetization switching.

This study not only provides evidence of direct control of magnetization in ferrimagnetic systems via molecular chirality, it highlights how combining chiral molecules with magnetic materials that have more complex ordering can result in unexpected phenomena. We anticipate this research will broaden the scope of magnetic materials used for CISS studies and open new avenues for the design of energy-efficient chiral spintronics devices.

## EXPERIMENTAL SECTION

*Chemicals and Materials*

Unless otherwise stated, chemicals were purchased from chemical suppliers and used without any additional purification. D-leucine (AK Scientific, 99%), L-leucine (AK Scientific, 99%), ammonia (aqueous) (Ajax, 28-30% w/w), copper nitrate hemi pentahydrate (ECP .Ltd, 99%), diethyl ether (Macron, 99%), ethanol (ECP .Ltd, 99.9%), methanol (JT Baker, 99.8%), triethylamine (JT Baker, >99%), were obtained from commercial sources and used without further purification.

*Physical Measurements*

Infrared spectroscopy experiments were collected using a Bruker Vertex 70 FT-IR/FT-FIR spectrometer. Raman spectra were collected using a Hiroba Labram HR evolution confocal Raman spectrometer at room temperature utilizing and air cooled 1064 nm 500 mW Yag laser or a 532 nm 100 mW 1 MHz air cooled Yag laser. Experiments were run using a 10× or 50× objective lens with either an 1800 gr/mm or an 1800 gr/mm and 600 gr/mm gratings in place inside of an 800 mm focal length flat field monochromator. Experiments were performed with varying exposure times and exposure events. A multichannel air-cooled array detector was used to collect spectra for non-fluorescent non-aromatic compounds and a liquid nitrogen cooled InGaAs line detector was used to collect spectra for fluorescent aromatic compounds.

Samples were dried after synthesis and prior to elemental analysis under vacuum ($< 4\times10^{-1}$ mbar). Elemental analysis measurements were conducted using a Vario EL cube (Elementar Analysensysteme GmbH, Germany) for the amino acid Cu(II) complexes.

*Synthesis of copper chiral materials*

The chiral copper coordination complexes were synthesized according to the literature procedure with slight modification.[16] An aqueous 0.1M triethylamine (TEA) solution was prepared by combining TEA (139 μL) and water (9.861 mL). Then L-leucine or D-leucine amino acid were added to the TEA solution, and was stirred overnight under nitrogen before $Cu(NO_3)_2 .2.5H_2O$ (1 eq.) dissolved in water (5 mL) was added dropwise to the amino acid solution causing the precipitation of a blue solid. The solid was collected via centrifugation. The precipitate was washed first with water (2 × 5 mL) then ethanol (2 × 5 mL) followed by washing with diethyl ether (2 × 5mL), centrifuging, and decanting between each wash. The precipitate was then air dried to yield a blue powder. Yield: 23% - 87%.

**[Cu(L-leucine)$_2$]:**

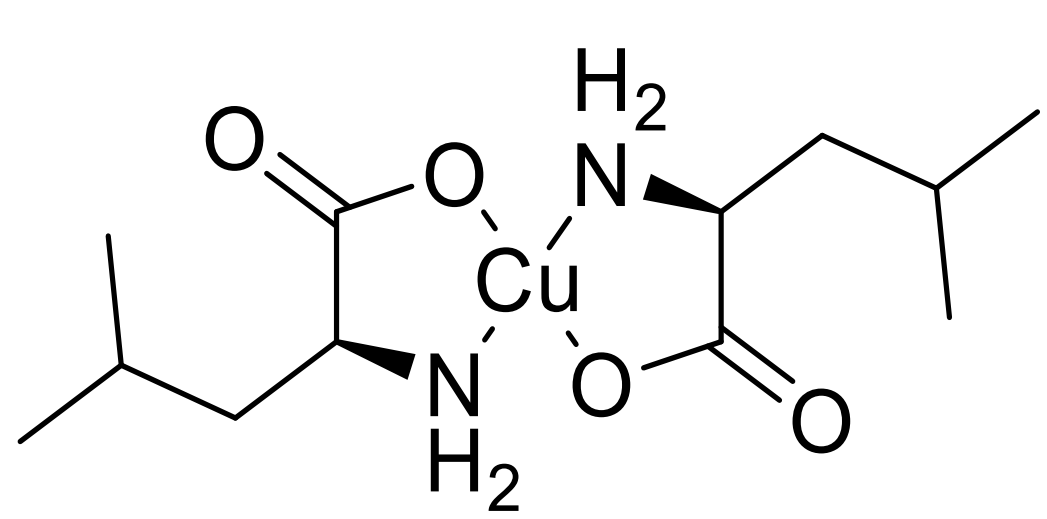

[Cu(L-leucine)$_2$] was synthesized according to the general procedure. L-leucine (1 mmol, 131 mg) and $Cu(NO_3)_2.2.5\ H_2O$ (0.5 mmol, 116 mg). Yield (light blue powder): 37 mg (23 %). Anal. Found: N, 8.84; C, 44.01; H, 7.27. Calcd for $C_{12}H_{24}CuN_2O_4$: N, 8.65; C, 44.5; H, 7.47.

IR (ATR): ν/cm$^{-1}$ = 3319*w* (NH$_2$), 3243*m* (NH$_2$), 2960*m* (C-H), 2926*w* (C-H), 2870*w* (C-H), 1614*vs* (C=O), 1566*m*, 1473*m*, 1333*m*, 1306*m*, 1107*s*, 852*m*, 788*m*, 688*s*, 569*s*, 437*m*. Raman (532 nm, 50x, 2%, 10s): cm$^{-1}$ = 3318*m* (N-H), 3243*m* (N-H), 2926*s* (C-H), 2869*s* (C-H), 2761*vw*, 2716*w*, 1667*w* (C=O), 1597*m* (NH$_2$), 1454*s*, 1309*m*, 1113*m*, 832*s* (C-COO-Cu), 562*s*, 433*m* (Cu-O), 246*w* (Cu-N).

**[Cu(D-leucine)$_2$]**:

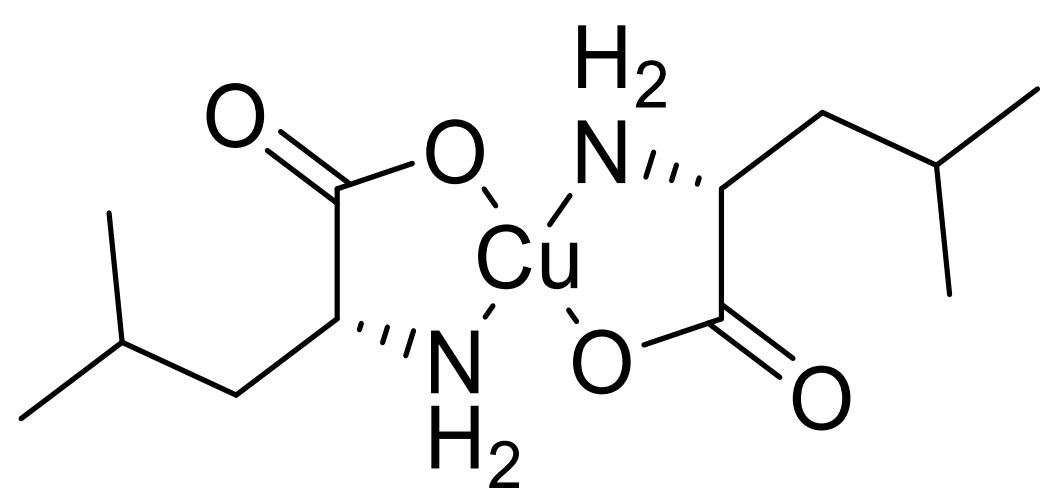

[Cu(D-leucine)$_2$] was synthesized according to the general procedure. D-leucine (1 mmol, 131 mg) and $Cu(NO_3)_2.2.5H_2O$ (0.5 mmol, 116 mg). Yield (Light blue powder): 136 mg (84%). Anal. Found: N, 8.79; C, 43.76; H, 7.24. Calcd for $C_{12}H_{24}CuN_2O_4$: N, 8.65; C, 44.5; H, 7.47.

IR (ATR): ν/cm$^{-1}$ = 3318*w* (NH$_2$), 3243*m* (NH$_2$), 2960*m* (C-H), 2926*w* (C-H), 2871*w* (C-H), 1614*vs* (C=O), 1566*m*, 1473*m*, 1333*m*, 1306*m*, 1107*s*, 853*m*, 788*m*, 688*s*, 570*s*, 438*m*.

*Thin films and sample preparation for MCD measurements*

TmIG thin films, 35 nm thick, were deposited on quartz substrates at room temperature using RF magnetron sputtering. The deposition utilized a $Tm_3Fe_5O_{12}$ target in an argon (Ar) atmosphere at a working pressure of $2 \times 10^{-2}$ Torr. To achieve a high vacuum environment, the base pressure in the sputtering chamber was reduced to below $1 \times 10^{-6}$ Torr using a turbomolecular pump. The sputtering was conducted at a power of 50 W. Initially, the as-sputtered TmIG films were amorphous and exhibited no detectable ferromagnetic properties. Subsequently, the TmIG thin films were annealed at 900°C for 5 hours in an air atmosphere to enhance their crystallinity.

MCD signals were measured at room temperature using a Jasco J-815 spectrometer equipped with a 450 W xenon lamp. White light was directed into a monochromator and then converted into linearly polarized light. This light was further modulated by a photo-elastic modulator operating at 50 kHz, converting the polarization from linear to circular. This modulation allowed simultaneous recording of MCD spectra using a photomultiplier for both right and left circularly polarized light. The incident light was aligned parallel to the applied magnetic field direction and normal to the sample plane. Background data, including contributions from the substrate, were subtracted to obtain absolute spectra of the samples. To prepare the experiment, we start by preparing a 1 mL solution of 10 mM L-leucine or D-leucine. The TmIG thin film was magnetized using a magnetic field of 0.8 T and then the field reduced to zero to set the remnant magnetization (0 T). The TmIG thin film was placed into the prepared L-leucine and D-leucine solutions separately and left in the solution for 1 day to dry. Finally, the CD spectra of the dried TmIG thin films was measured using a large area spot size of 4 mm x 4 mm.

*X-ray Absorption Near Edge Spectroscopy (XANES) measurements and simulations*

The XANES data was measured by the Lytle detector at Taiwan beamline BL12B2, SPring-8. The beamline resolving power is ~10000, the angle between the beam and detector is 90˚, and the incident angle is ~45˚ to the normal direction of the sample surface. The measurement was done without applying a magnetic field at room temperature and in the air atmosphere.

The XANES simulation was performed with FDMNES code using the known crystal structures,[27, 28] the lattice constant 12.324 Å, and the Ia-3d space group. The radius R= 7 Å of the cluster is used for the self-consistent calculations. We set the initial condition of ferrimagnetism as the electron configuration on Fe($O_h$) and Fe($T_d$) sites without including the spin status on the Tm site. The relativistic spin-orbital effect is included in the simulation due to the heavy-element Tm in the system. Because of the region of interest in XANES, the multiple scattering mode with muffin-tin approximation is used in the code.

*Computational Methods*

Calculations were carried out using Gaussian16 rev. C.01"Gaussian 16, Revision C.01." and visualised with GaussView6.[29] The M06-2X functional[30] with the 6-311+G(d,p) basis set was employed for geometry optimisations. The solution-phase was modelled by the PCM solvation model using water as the solvent.[31] The calculations are all open-shell due to the unpaired electron in the d-AO manifold requiring a doublet multiplicity. An ultrafine grid (pruned grid with 99 radial shells and 590 angular points per shell) was specified The SCF convergence criterion was set to $10^{-9}$, which equates to $<10^{-9}$ RMS change in the density matrix and $<10^{-7}$ maximum change in the density matrix. All structures were optimised under no symmetry constraints and confirmed as minima by the presence of no imaginary frequencies.

The initial geometries of both Cu(II) complexes of l- and d-leucine have been taken from X-ray crystallography data for structurally related molecules reported in the literature[16,24]. The structures were optimised and confirmed as minima at the M06-2X-PCM(water)/6-311+G(d,p) level of theory. A conformational survey was carried out with CREST[32] (employing default settings), and the lowest energy conformers were subsequently optimised at the M06-2X-PCM(water)/6-311+G(d,p) level. Due to substituent orientation and variations in H-bonding a range of conformers of both the d- and l-Phenylalanine-Cu complexes is possible, however, the complexes evaluated here are expected to be representative.

ECD calculations require the use of a functional with improved long-range exchange, and the CAM-B3LYP functional is known to provide good results.[33,34] ECD calculations have been carried out as single-point computations on the optimised solution-phase M06-2X-PCM(water)/6-311+G(d,p) geometries at the TDDFT-CAM-B3LYP-PCM(water)/6-311+G(d,p) level. 120 excited states (of doublet character) were included. The ECD spectra were extracted from GaussView6 and compared with experimental spectra, a systematic shift in the wavelength and intensity was made to visually align computational with experimental spectra. A qualitative visual match has been obtained (Fig. S2 and S3).

**Supporting Information**

The Supporting Information is available free of charge.

CD spectra of TmIG before and after the adsorption of different Leucine molecular.

Comparison of experimental and calculated CD spectra of the Cu(II) complex of D-leucine and L-leucine.

**Author Contributions**

W.-H. Liao and J. S. Webb contributed equally to this work.

**Competing Financial interests**

The authors declare no competing financial interest.

**ACKNOWLEDGMENTS**

This work was supported by the National Science and Technology Council, Taiwan, for financially supporting this research under Contract No. MOST 111-2112-M-153 -002 -MY3 and 112-2918-I-153-001, and NSTC 113-2112-M-213 -025 -MY3. This research was also supported in part by Higher Education Sprout Project, Ministry of Education to the Headquarters of University Advancement at National Cheng Kung University (NCKU). J.W. thanks the University of Auckland for a Doctoral Scholarship. The MacDiarmid Institute for Advanced Materials and Nanotechnology is supported under the New Zealand Centers of Research Excellence Program.

We would like to express our heartfelt gratitude to the late Prof. Jiunn Chen for his invaluable discussions on the electronic structure of ferrimagnetic oxides and MCD, which have greatly inspired this work. His past contributions to the field continue to be a guiding influence on our research, and we commemorate his enduring impact on this study.